\documentclass[a4paper]{revtex4}
\usepackage{graphicx}
\usepackage{fancyhdr}
\usepackage{amsmath}
\usepackage{color}

\begin{document}

\title{Constraints on the Georgi-Machacek Model by Current LHC Data}


\author{Cheng-Wei Chiang}
\affiliation{Center for Mathematics and Theoretical Physics and Department of Physics, 
National Central University, Chungli, Taiwan 32001, R.O.C.}
\affiliation{Institute of Physics, Academia Sinica, Taipei, Taiwan 11529, R.O.C.}
\affiliation{Physics Division, National Center for Theoretical Sciences, 
Hsinchu, Taiwan 30013, R.O.C.}

\begin{abstract}
In the Georgi-Machacek model with a custodial symmetry in the Higgs potential and vacuum alignment, the triplet vacuum expectation value is allowed to be of ${\cal O}(10)$ GeV, which leads to the possibility of significant modifications in the couplings of the SM-like Higgs bosons $h$ with other SM particles.  In this talk given at the HPNP2015 conference held at Toyama University, we review constraints on the model based on the latest LHC data of the SM-like Higgs boson, like-sign $W$ boson events, and searches for additional neutral Higgs bosons.  In particular, we concentrate on the parameter space for small mixing angle $\alpha$ between the two custodial singlets.  It is pointed out that constraints from the non-SM custodial singlet are most constraining and those from the 5-plet are independent of $\alpha$.  While currently there is no constraint from the 3-plet, we show that its $f \bar f$ and $\gamma\gamma$ channels through the gluon fusion production can be very promising for searches or constraining in the mass range between 160 GeV and 350 GeV because of its gauge-phobic property.
\end{abstract}

\maketitle

\thispagestyle{fancy}


%%%%%%%%%%%%%%%%%%%%%%%%%%%%%%%%%%
\section{Introduction}
%%%%%%%%%%%%%%%%%%%%%%%%%%%%%%%%%%

Even though the 125-GeV boson is found to have many properties very similar to the Higgs boson in the standard model (SM), it is still far from clear whether it is the sole entity responsible for the breakdown of electroweak symmetry and mass of all elementary particles.  Besides, there is no established guiding principle about the structure of the Higgs sector other than the Lorentz and gauge symmetries.  It is therefore not unnatural to consider additional Higgs representations that also contribute to the symmetry breaking and may have a connection to a hidden sector.

As a new physics model with an extended Higgs sector, the Georgi-Machacek (GM) model~\cite{Georgi:1985nv,Chanowitz:1985ug} has some intriguing features not shared by commonly considered models whose extra Higgs fields are only singlets and/or doublets.  In addition to a doublet field $\phi$ with $Y = 1/2$ as in the SM, the GM model has a triplet field $\Delta$ composed of a complex triplet $\chi$ of hypercharge $Y = 1$ and a real triplet $\xi$ of $Y = 0$ under the SM $SU(2)_L \times U(1)_Y$ gauge symmetry.  Starting with a Higgs potential with custodial symmetry and vacuum alignment between the complex and real triplets, the model preserves the electroweak rho parameter $\rho = 1$ at tree level.  This allows the possibility of the triplet vacuum expectation value (VEV), $v_\Delta$, as large as up to a few tens of GeV.  The model has many Higgs bosons, including the SM-like Higgs boson $h$ and another singlet $H_1$, one 3-plet $H_3$, and one 5-plet $H_5$, with mass degeneracy within each multiplet as a result of the custodial symmetry~\cite{Gunion:1989ci,Chiang:2012cn}.  Due to the mixing between the Higgs doublet and triplet fields, the couplings between the SM-like Higgs boson and the weak gauge bosons can be stronger than their SM values~\cite{Logan:2010en,Falkowski:2012vh,Chang:2012gn,Chiang:2013rua}, leading to interesting collider phenomenology~\cite{Godfrey:2010qb,Chiang:2012dk,Englert:2013zpa,Chiang:2013rua}.  In the case of large triplet VEV, the 5-plet couples dominantly to the weak gauge bosons.  Therefore, vector boson fusion processes serve the most promising channels to search for such exotic Higgs bosons and verify their mass degeneracy at the LHC~\cite{Chiang:2013rua}.  Besides, the model has the $H_3^{\pm} W^\mp Z$ vertex at tree level.  Such a vertex is known to be small in multidoublet models, because they appear only at loop levels.  Besides, neutrinos can obtain Majorana mass from the Higgs triplet VEV through the so-called type-II seesaw mechanism.  The couplings between the triplet field and leptons lead to lepton number-violating processes and possibly even lepton flavour-violating ones.

Many of the above-mentioned properties and couplings depend on the value of the triplet VEV, $v_\Delta$, which serves as a quantitative indicator for the participation of the Higgs triplet in the electroweak symmetry breaking.  It is therefore of great interest to experimentally determine or constrain this parameter in the model.  In the following, we discuss how some of the LHC data have been used to put constraints on the model, particularly in the scenario where the mixing between the doublet and the triplet is small, as favoured by the SM-like Higgs data.

%%%%%%%%%%%%%%%%%%%%%%%%%%%%%%%%%%
\section{Basics of Georgi-Machacek Model}
%%%%%%%%%%%%%%%%%%%%%%%%%%%%%%%%%%

It is more convenient to organise the isospin doublet field $\phi$ and the triplet fields
 $\chi$ with $Y=1$ and $\xi$ with $Y=0$~\footnote{The convention is such that the electric charge $Q = T_3 + Y$ with $T_3$ being the third isospin number.} in an $SU(2)_L\times SU(2)_R$ covariant form:
\begin{align}
\Phi=
\begin{pmatrix}
\phi^{0*} & \phi^+ \\
-(\phi^+)^* & \phi^0
\end{pmatrix} ~,~
\Delta=
\begin{pmatrix}
\chi^{0*} & \xi^+ & \chi^{++} \\
-(\chi^+)^* & \xi^0 & \chi^{+} \\
(\chi^{++})^* & -(\xi^+)^* & \chi^{0} 
\end{pmatrix} ~,
\label{eq:Higgs_matrices}
\end{align}
where the phase convention for the component scalar fields is such that $\phi^- = (\phi^+)^*$, $\chi^{--} = (\chi^{++})^*$, $\chi^{-} = (\chi^{+})^*$, $\xi^{-} = (\xi^{+})^*$.  Moreover, the neutral components after electroweak symmetry breaking are parameterised as 
\begin{align}
\phi^0 = \frac{1}{\sqrt{2}}(v_\phi+\phi_r+i\phi_i) ~,~ 
\chi^0 = v_\chi+\frac{1}{\sqrt{2}}(\chi_r+i\chi_i) ~,~
\xi^0 = v_\xi+\xi_r ~, \label{eq:neutral}
\end{align}
where $v_\phi$, $v_\chi$ and $v_\xi$ denote the VEV's of $\phi$, $\chi$ and $\xi$, respectively.  
The explicit form of most general Higgs potential consistent with the $SU(2)_L\times SU(2)_R\times U(1)_Y$ symmetry can be found, for example, in Ref.~\cite{Chiang:2012cn}.  The physical 5-plet, $H_5 = (H_5^{\pm\pm},H_5^\pm,H_5^0)^T$, arises within the $\Delta$ field.  The two 3-plet fields mix through the angle $\beta$ to render a physical CP-odd Higgs 3-plet, denoted by $H_3 = (H_3^\pm,H_3^0)^T$, and another NG 3-plet, $(G^\pm,G^0)^T$, to become the longitudinal components of the weak gauge bosons.  The two CP-even singlet fields further mix by an angle $\alpha$, determined by the quartic coupling constants in the Higgs potential, to produce the SM-like Higgs boson $h$ and another physical singlet denoted by $H_1^0$.

Under the vacuum alignment assumption $v_\chi=v_\xi \equiv v_\Delta$, the masses of the $W$ and $Z$ bosons have exactly the same form as in the SM: $M_W^2 = \frac{g^2v^2}{4}$ and $M_Z^2=\frac{g^2v^2}{4\cos^2\theta_W}$, where $\theta_W$ is the weak mixing angle and $v^2\equiv v_\phi^2 + 8v_3^2 = (246~{\rm GeV})^2$.  Therefore, the electroweak rho parameter is unity at tree level.  Define the ratio of the VEV's as
\begin{align}
\tan\beta = \frac{v_\phi}{2\sqrt{2} v_\Delta} ~,
\end{align}
which is the reciprocal of $\tan\theta_H$ used in most other works and goes to infinity in the SM limit.  If one requires that the fermion mass comes from Yukawa couplings with the Higgs doublet and the top Yukawa coupling is perturbative, then the triplet VEV has an upper bound $v_\Delta \lesssim 80$ GeV.

%%%%%%%%%%%%%%%%%%%%%%%%%%%%%%%%%%
\section{Experimental Constraints}
%%%%%%%%%%%%%%%%%%%%%%%%%%%%%%%%%%

Recently, the same-sign diboson process $pp \to W^\pm W^\pm jj$ had been measured at the LHC using leptonic decay
channels of $W$ bosons, with production cross sections of two fiducial regions reported to be consistent with
the Standard Model expectations within $1\sigma$~\cite{ATLAS2014}.  The results can be used to constrain new physics models with a modified quartic $W$ vertex, as in the case of the GM model due to the mediations of exotic Higgs bosons.

%%%%%%%%%%%%%%%%%%%%%%%%%%%%%%%%%%
\begin{figure}[ht]
\bigskip
\bigskip
\centering
\includegraphics[width=100mm]{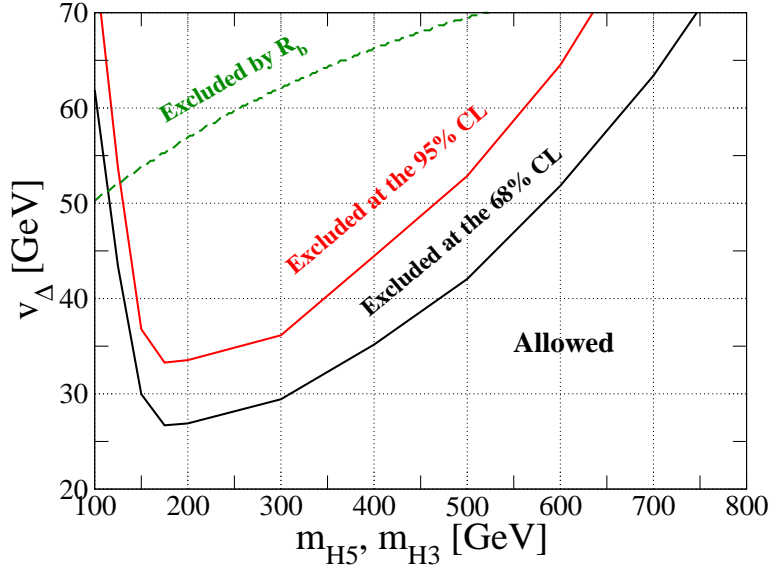}
\caption{Constraint on the $m_{H_5}$-$v_\Delta$ plane by ATLAS data at 68\% and 95\% CL (reproduced from Ref~\cite{Chiang:2014bia}).}
\label{fig:H5cc}
\end{figure}
%%%%%%%%%%%%%%%%%%%%%%%%%%%%%%%%%%

Fig.~\ref{fig:H5cc} shows the constraint on the $m_{H_5}$-$v_\Delta$ plane by the 20.3 fb$^{-1}$ data of 8-TeV LHC provided by the ATLAS Collaboration.  The region above the black (red) curve is excluded at the 68\% (95\%) confidence level (CL).  The most severe upper bound on $v_\Delta$ is about 33 GeV for $m_{H_5} = 200$ GeV.  The bound becomes weaker as $m_{H_5}$ becomes larger and approaches the above-mentioned 80 GeV bound at $m_{H_5} \sim 700$ GeV.
In the case of $m_{H_5} < 200$ GeV, a milder bound on $v_\Delta$ is also obtained, as more events from the
$H_5^0$ contribution are rejected by the kinematic cuts used by ATLAS.  At the 14-TeV LHC, the discovery reach becomes the largest also at around $m_{H_5} = 200$ GeV, where a $5\sigma$ discrepancy is expected in the cases of $v_\Delta
\agt 24$, 17, 12 and 7 GeV for the luminosity of 30, 100, 300, and 3000 fb$^{-1}$, respectively.

%%%%%%%%%%%%%%%%%%%%%%%%%%%%%%%%%%
\begin{table}
\caption{Couplings of the neutral Higgs bosons to SM quarks and weak gauge bosons in units of their SM values.  $\eta_f = +1$ for up-type quarks and $-1$ for down-type quarks and charged leptons.}
\medskip
\renewcommand{\arraystretch}{1.2}
\begin{tabular}{c|ccc}
\hline\hline
Higgs & ~~~~$\kappa_F$~~~~ & ~~~~
& ~~~~~~~~~~~~~~~~~$\kappa_V$~~~~~~~~~~~~~~~~~ \\
\hline
$h$ & $\displaystyle \frac{\cos\alpha}{\sin\beta}$
&  
& $\displaystyle \sin\beta\cos\alpha - \sqrt{\frac83} \cos\beta\sin\alpha$
\\
$H_1^0$ & $\displaystyle \frac{\sin\alpha}{\sin\beta}$ 
&
& $\displaystyle \sin\beta\sin\alpha + \sqrt{\frac83} \cos\beta\cos\alpha$
\\
$H_3^0$ & $i\eta_f\cot\beta$ 
& & 0
\\
$H_5^0$ & 0 
&
& $\displaystyle \kappa_W = -\frac{\cos\beta}{\sqrt{3}}$ and 
$\displaystyle \kappa_Z = \frac{2\cos\beta}{\sqrt{3}}$
\\
\hline\hline
\end{tabular}
\label{tab:couplings}
\end{table}
%%%%%%%%%%%%%%%%%%%%%%%%%%%%%%%%%%

After the discovery of the 125-GeV Higgs boson, efforts have been made to search for another neutral Higgs boson through different channels over a wide range of mass.  Such results can also be used to impose constraints on the GM model.  In Table~\ref{tab:couplings}, we list the couplings of the neutral Higgs bosons in the model to the SM fermions and weak gauge bosons, all normalised to their SM values.  Because of the factor of $\sqrt{\frac83}$ in $\kappa_V$ of $h$ and $H_1^0$, their values can be larger than 1 with a maximum of $\sqrt{\frac83}$.  In the case of $H_5^0$, $\kappa_Z$ is larger than $\kappa_W$ in magnitude by a factor of 2.  As a result, $Br(ZZ)$ branching ratio will be larger than $Br(WW)$ by about a factor of 2 in the high-mass regime of $H_5^0$.  This is a nice discriminant for the neutral Higgs boson originated from the custodial 5-plet.  $H_3^0$ is gauge-phobic, while $H_5^0$ is quark-phobic.  In the small $\alpha$ limit, $\kappa_F^{H_1} \sim \frac{\alpha}{\sin\beta}$ and $H_1^0$ becomes more fermiophobic.

In the assumption of narrow width for $\varphi$ ($\varphi = h, H_1^0, H_3^0, H_5^0$), we define the signal strength
\begin{align}
\mu_{X}[\varphi] = 
\frac{\sigma^{\rm GM}(pp \to \varphi)~{\cal B}^{\rm GM}(\varphi \to X)}
     {\sigma^{\rm SM}(pp \to \varphi)~{\cal B}^{\rm SM}(\varphi \to X)}
\end{align}
where $X$ denotes some decay mode of $\varphi$.  By incorporating the scaling factors given in Table~\ref{tab:couplings} for $h$ and fixing its mass at 125 GeV, one can perform a global fit to the measured signal strengths for the SM-like Higgs boson.  Using the latest data~\cite{Khachatryan:2014jba,ATLAS2015}, we obtain $-20^\circ \lesssim \alpha \lesssim 0^\circ$ by taking the heavy exotic Higgs masses limit.  We therefore concentrate on the examples of $\alpha = 0$, $-\pi/24$, and $-\pi/12$ in the following analyses.

\begin{figure}[ht]
\centering
\includegraphics[width=55mm]{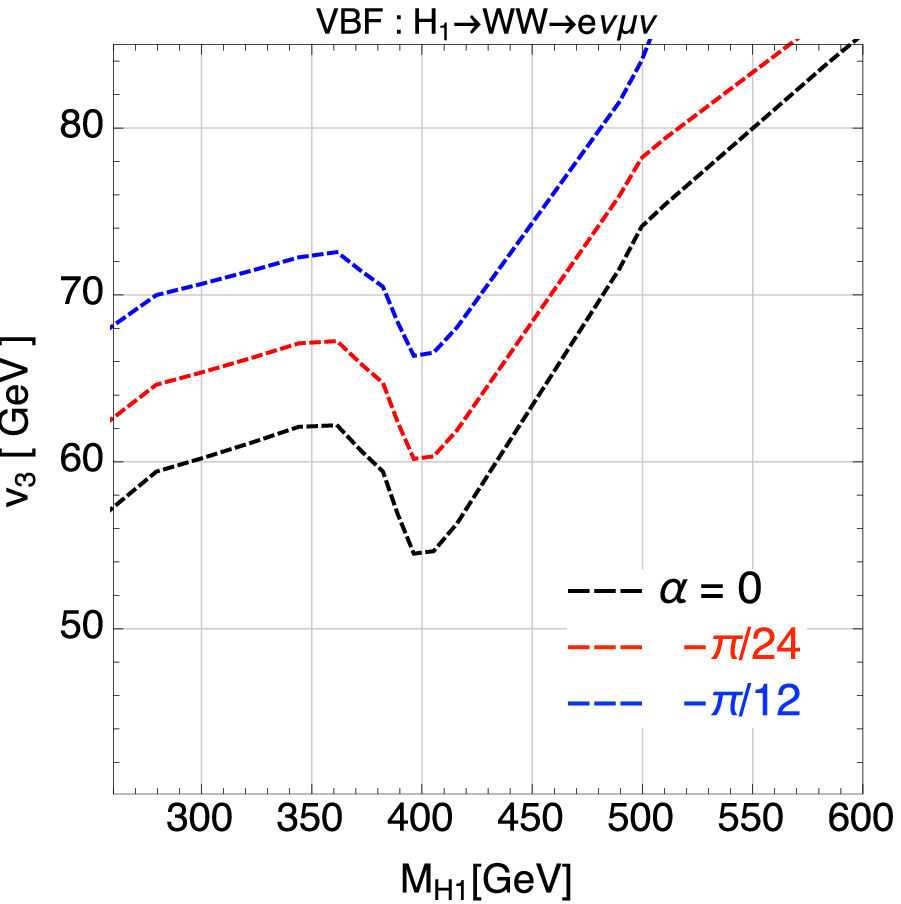}
\includegraphics[width=55mm]{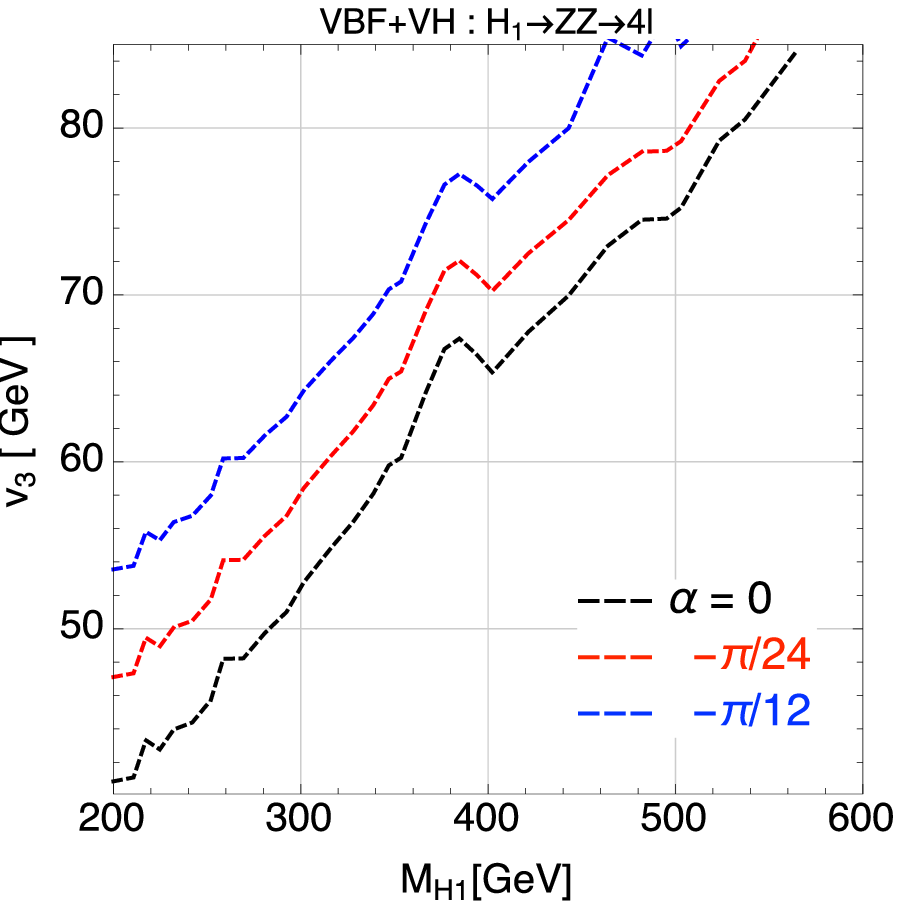}
\includegraphics[width=55mm]{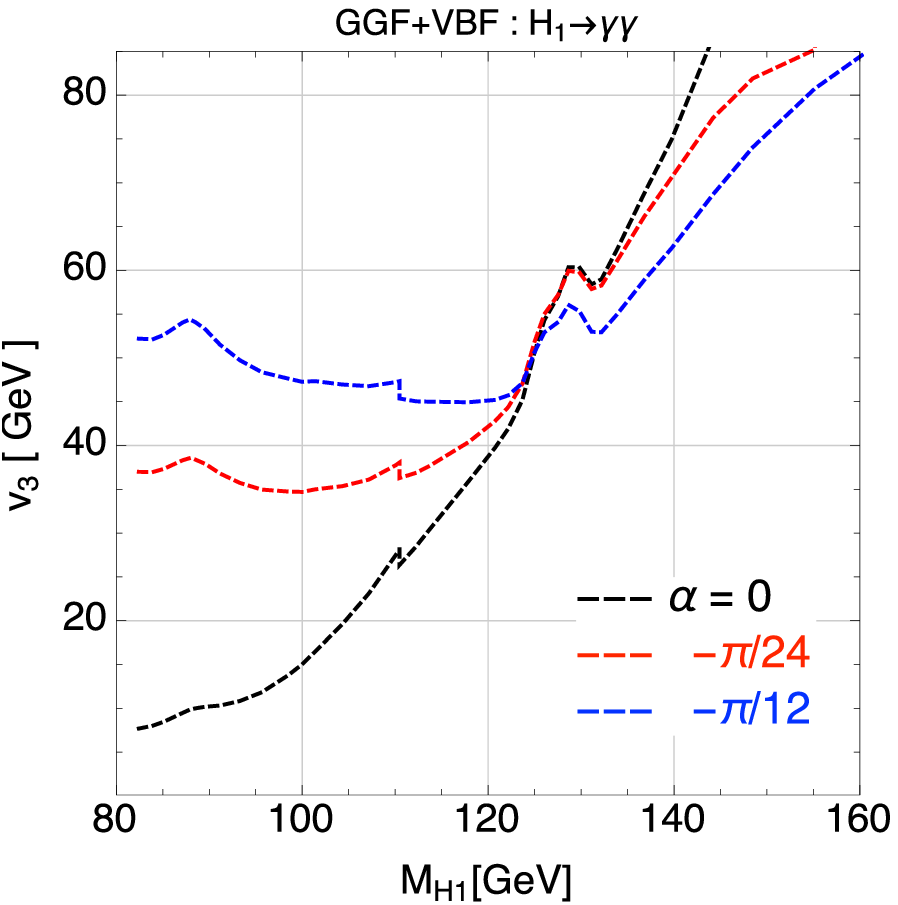}
\\
(a) \hspace{50mm} (b) \hspace{50mm} (c)
\caption{Upper limits on $v_\Delta (= v_3)$ as a function of $M_{H_1}$ for the (a) $WW$, (b) $ZZ$, and (c) $\gamma\gamma$ channels through the VBF mechanisms. (reproduced from Ref.~\cite{Chiang:2015kka}).}
\label{fig:H1}
\end{figure}

In the case of the other exotic Higgs bosons in the model, their masses are unfixed parameters.  Therefore, one has for $H_1^0$ as an example
\begin{align}
\mu_X^\text{GGF} [H_1]
&=
(\kappa_F^{H_1})^2 \times
\frac{{\mathcal B}_X}{{\mathcal B}_X^\text{SM}(M_{H_1})} 
\simeq 
\frac{(\kappa_F^{H_1})^2(\kappa_X^{H_1})^2}
{(\kappa_V^{H_1})^2{\mathcal B}_V^\text{SM}(M_{H_1})
+ (\kappa_F^{H_1})^2{\mathcal B}_F^\text{SM}(M_{H_1})} ~,
\notag \\
\mu_X^\text{VBF} [H_1]
&=
(\kappa_V^{H_1})^2 \times
\frac{{\mathcal B}_X}{{\mathcal B}_X^\text{SM}(M_{H_1})} 
\simeq 
\frac{(\kappa_V^{H_1})^2(\kappa_X^{H_1})^2}
{(\kappa_V^{H_1})^2{\mathcal B}_V^\text{SM}(M_{H_1})
+ (\kappa_F^{H_1})^2{\mathcal B}_F^\text{SM}(M_{H_1})} ~,
\end{align}
where ${\mathcal B}_V^\text{SM}(M_{H_1})$ and ${\mathcal B}_F^\text{SM}(M_{H_1})$ denote the inclusive branching ratios of a SM Higgs of fictitious mass $M_{H_1}$ decaying into a pair of vector bosons and fermions, respectively, with all other modes ({\it e.g.}, $\gamma\gamma$, $Z\gamma$ and multi-particles) neglected in last expressions.  The superscript GGF means the production channel, including $ggh$ and $t\bar t h$ processes, and VBF includes the vector boson fusion and associated productions.  As a result of suppression in the coupling with fermions, the GGF production of $H_1^0$ is significantly smaller than the VBF process.  Therefore, the VBF search channels impose stronger constraints on the parameter space, as shown in Fig.~\ref{fig:H1}.  Comparing the plots, one notices that in the higher mass regime the $ZZ$ channel is generally more constraining than the $WW$ channel except for the region 375 GeV $\alt M_{H_1} \alt 450$ GeV, in which the former (latter) has a slightly worse (better) sensitivity experimentally.  The $\gamma\gamma$ channel has more constraining power in the low-mass regime.  Plot (c) also shows the change in the interference pattern at $M_{H_1} = 125$ GeV for different choices of $\alpha$.

%%%%%%%%%%%%%%%%%%%%%%%%%%%%%%%%%%
%%%%%%%%%%%%%%%%%%%%%%%%%%%%%%%%%%

It is noted that the $H_1^0 \to hh$ decay mode is not included in the calculations, because the relevant coupling, $λ_{hhH_1}$, is not determined without explicitly specifying the entire Higgs potential, thereby inevitably introducing uncertainties.  Qualitatively, a nonzero $λ_{hhH_1}$ coupling will result in a suppression of the regular search channels when $M_{H_1} \agt 250$ GeV, and may lead to more production of four-body final states.  Therefore, the bound for $M_{H_1} \agt 250$ GeV will generally become weaker.

%%%%%%%%%%%%%%%%%%%%%%%%%%%%%%%%%%
\begin{figure}[ht]
\centering
\includegraphics[width=70mm]{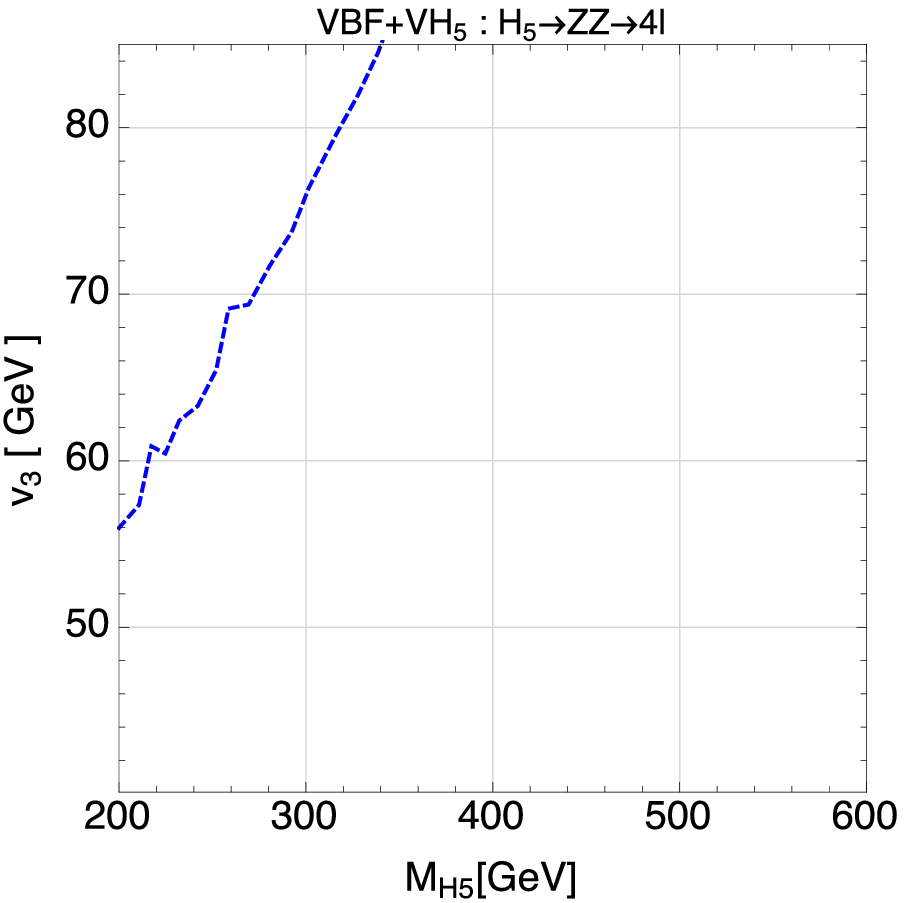}
\includegraphics[width=70mm]{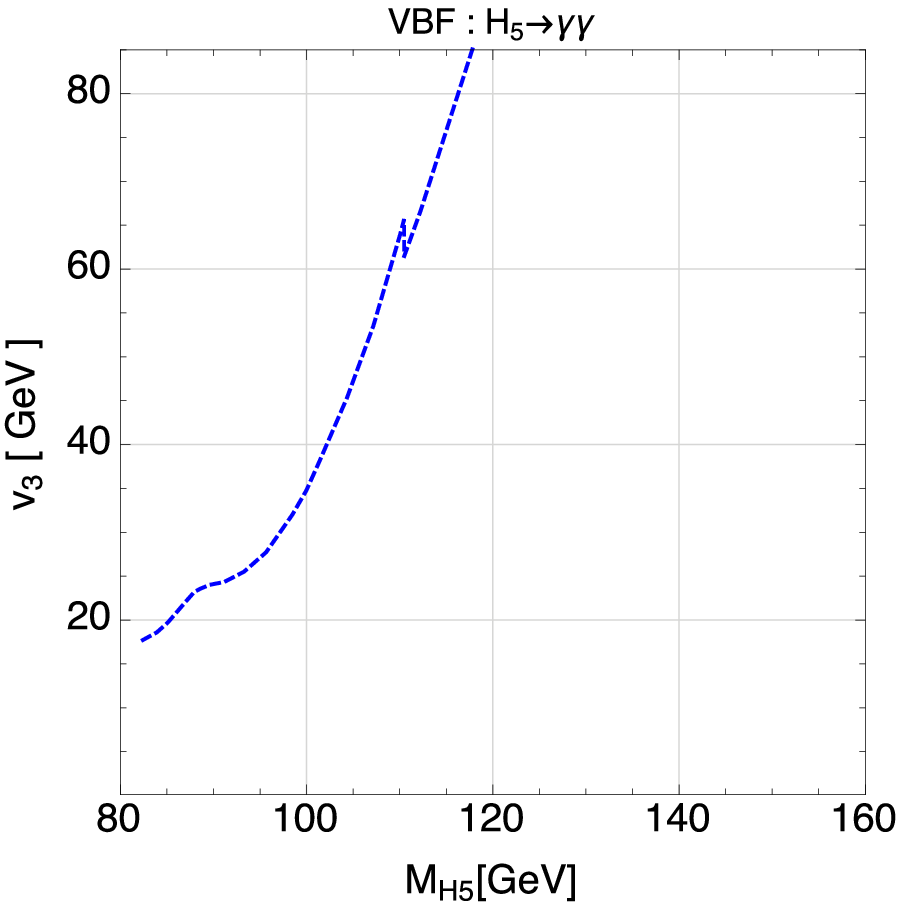}
\\
(a) \hspace{65mm} (b)
\caption{Upper limits on $v_\Delta$ as a function of $M_{H_5}$ for the (a) $ZZ$ and (b) $\gamma\gamma$ channels through the VBF mechanism. (reproduced from Ref.~\cite{Chiang:2015kka}).}
\label{fig:H5}
\end{figure}
%%%%%%%%%%%%%%%%%%%%%%%%%%%%%%%%%%

Similarly, one may also put an upper limit of the triplet VEV by considering $H_5^0$ as the intermediate Higgs boson.  Fig.~\ref{fig:H5} shows the bound as a function of $M_{H_5}$ from both $ZZ$ and $\gamma\gamma$ decay modes via the VBF mechanism.  Clearly, this constraint from the $H_5^0$ search is weaker than those presented in Fig.~\ref{fig:H1}.
This is related to the fact that the signal strength in this case is mainly enhanced in the low-mass regime only.  However, the bounds from $H_5^0$ is useful in the sense that unlike Fig.~\ref{fig:H1} for $H_1^0$, they are independent of the mixing angle $\alpha$.  We note in passing that no useful constraint can be obtained from the $WW$ mode yet, is a result of the non-universal scaling behaviors in the couplings with the weak bosons.

\begin{figure}[ht]
\centering
\includegraphics[width=70mm]{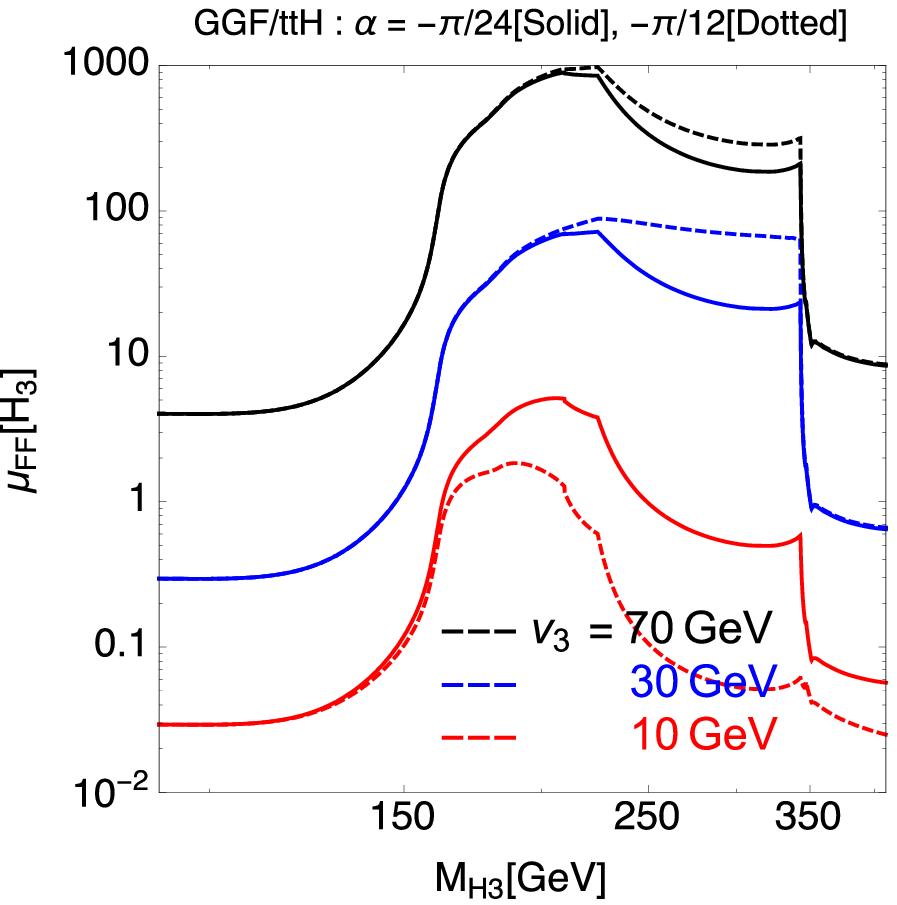}
\includegraphics[width=70mm]{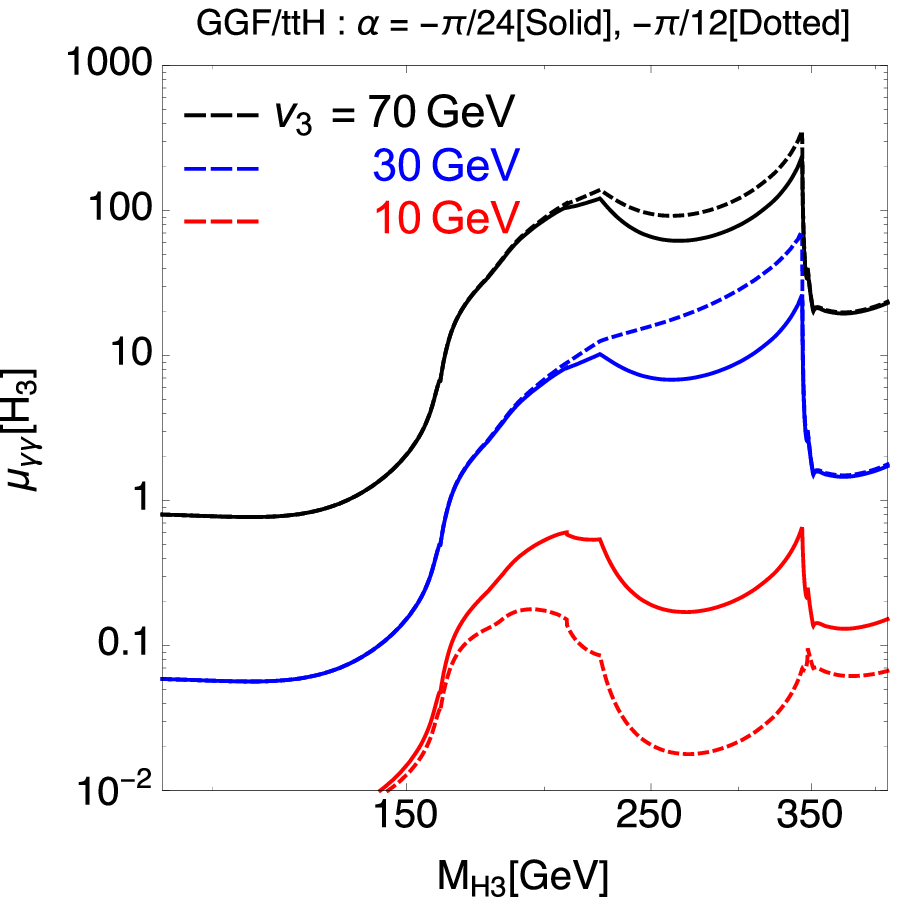}
\\
(a) \hspace{65mm} (b)
\caption{Signal strengths of the $H_3^0$ boson in the GGF production of the (a) $f \bar f$ and (b) $\gamma\gamma$
channels as functions of $M_{H_3}$. (reproduced from Ref.~\cite{Chiang:2015kka}).}
\label{fig:H3}
\end{figure}

Since the $H_3^0$ does not couple to the weak gauge bosons, one can only make use of the $f \bar f$ and $\gamma\gamma$ modes through the GGF production mechanism to search for the particle.  Moreover, $H_3^0$ is a CP-odd particle.  Therefore, the signal strengths for the fermion pair decays in the Born approximation are
\begin{align}
\mu_{FF}^\text{GGF} [H_3]
&=
(\kappa_F^{H_3})^2 
\frac{F_{1/2}^A(M_{H_3})}{F_{1/2}^S(M_{H_3})}
\times
\frac{{\mathcal B}_{F}}{{\mathcal B}_{F}^\text{SM}(M_{H_3})}
\left( 1 - \frac{4M_f^2}{M_{H_3}^2} \right)^{-1} ~,
\end{align}
where $F_{1/2}^S(M)$ and $F_{1/2}^A(M)$ are the fermionic loop functions for CP-even and -odd scalar
bosons, respectively.  Since the decays of a fictitious SM Higgs boson in the $M_{H_3} \alt 2M_W$ region are dominated by the fermion pairs, $\mu^{\rm GGF}_X [H_3] \sim \cot2\beta$.  On the other hand, when $M_{H_3} > 2M_W$, the inclusive BSM ${\mathcal B}_{F}^\text{SM}(M_{H_3})$ is very small.  Therefore, the signal strengths of the fermionic modes have an enhancement onset at around $2M_W$.  This enhancement is slightly reduced when the $hZ$ mode opens up and further reduced above the $t \bar t$ threshold, as shown in Fig.~\ref{fig:H3}.  Therefore, a search of such channels in the regime $2M_W \alt M_{H_3} \alt 2M_t$ can readily discover $H_3^0$ or put stringent constraints on the model parameters.

%%%%%%%%%%%%%%%%%%%%%%%%%%%%%%%%%%
%%%%%%%%%%%%%%%%%%%%%%%%%%%%%%%%%%

%%%%%%%%%%%%%%%%%%%%%%%%%%%%%%%%%%
\section{Conclusions}
%%%%%%%%%%%%%%%%%%%%%%%%%%%%%%%%%%

Because of the built-in custodial symmetry, the Georgi-Machacek model allows an ${\cal O}(10)$ GeV triplet vacuum expectation value.  Moreover, it offers the possibility of enhanced $hVV$ couplings than the standard model expectation.  The latest SM-like Higgs data favours the scheme where the mixing between the SM-like Higgs and the exotic singlet is small.  Using other search data at the LHC, we have put constraints on the parameter space (triplet vacuum expectation value versus exotic Higgs mass) of the model.

\begin{acknowledgments}
I would like to thank Shinya Kanemura and all the other organizers of HPNP2015 for the nice organization and great hospitality during the meeting.  This research was supported in part by the Ministry of Science and Technology of Taiwan under Grant No. NSC 100-2628-M-008-003-MY4.
\end{acknowledgments}

\bigskip % extra skip inserted

\begin{thebibliography}{99} % Use for 10-99 references

% original papers for the GM model
%\cite{Georgi:1985nv}
\bibitem{Georgi:1985nv} 
  H.~Georgi and M.~Machacek,
  %``Doubly Charged Higgs Bosons,''
  Nucl.\ Phys.\ B {\bf 262}, 463 (1985).
  %%CITATION = NUPHA,B262,463;%%
  
%\cite{Chanowitz:1985ug}
\bibitem{Chanowitz:1985ug} 
  M.~S.~Chanowitz and M.~Golden,
  %``Higgs Boson Triplets With M ($W$) = M ($Z$) $\cos \theta \omega$,''
  Phys.\ Lett.\ B {\bf 165}, 105 (1985).
  %%CITATION = PHLTA,B165,105;%%

% group classification
%\cite{Gunion:1989ci}
\bibitem{Gunion:1989ci} 
  J.~F.~Gunion, R.~Vega and J.~Wudka,
  %``Higgs triplets in the standard model,''
  Phys.\ Rev.\ D {\bf 42}, 1673 (1990).
  %%CITATION = PHRVA,D42,1673;%%

%\cite{Chiang:2012cn}
\bibitem{Chiang:2012cn} 
  C.~W.~Chiang and K.~Yagyu,
  %``Testing the custodial symmetry in the Higgs sector of the Georgi-Machacek model,''
  JHEP {\bf 1301}, 026 (2013)
  [arXiv:1211.2658 [hep-ph]].
  %%CITATION = ARXIV:1211.2658;%%

% large hVV couplings
%\cite{Logan:2010en}
\bibitem{Logan:2010en} 
  H.~E.~Logan and M.~A.~Roy,
  %``Higgs couplings in a model with triplets,''
  Phys.\ Rev.\ D {\bf 82}, 115011 (2010)
  [arXiv:1008.4869 [hep-ph]].
  %%CITATION = ARXIV:1008.4869;%%

%\cite{Falkowski:2012vh}
\bibitem{Falkowski:2012vh} 
  A.~Falkowski, S.~Rychkov and A.~Urbano,
  %``What if the Higgs couplings to W and Z bosons are larger than in the Standard Model?,''
  JHEP {\bf 1204}, 073 (2012)
  [arXiv:1202.1532 [hep-ph]].
  %%CITATION = ARXIV:1202.1532;%%

%\cite{Chang:2012gn}
\bibitem{Chang:2012gn} 
  S.~Chang, C.~A.~Newby, N.~Raj and C.~Wanotayaroj,
  %``Revisiting Theories with Enhanced Higgs Couplings to Weak Gauge Bosons,''
  Phys.\ Rev.\ D {\bf 86}, 095015 (2012)
  [arXiv:1207.0493 [hep-ph]].
  %%CITATION = ARXIV:1207.0493;%%
  
\bibitem{Chiang:2013rua} 
  C.~W.~Chiang, A.~L.~Kuo and K.~Yagyu,
  %``Enhancements of weak gauge boson scattering processes at the CERN LHC,''
  JHEP {\bf 1310}, 072 (2013)
  [arXiv:1307.7526 [hep-ph]].
  %%CITATION = ARXIV:1307.7526;%%

% pheno studies
%\cite{Godfrey:2010qb}
\bibitem{Godfrey:2010qb} 
  S.~Godfrey and K.~Moats,
  %``Exploring Higgs Triplet Models via Vector Boson Scattering at the LHC,''
  Phys.\ Rev.\ D {\bf 81}, 075026 (2010)
  [arXiv:1003.3033 [hep-ph]].
  %%CITATION = ARXIV:1003.3033;%%  

%\cite{Chiang:2012dk}
\bibitem{Chiang:2012dk} 
  C.~W.~Chiang, T.~Nomura and K.~Tsumura,
  %``Search for doubly charged Higgs bosons using the same-sign diboson mode at the LHC,''
  Phys.\ Rev.\ D {\bf 85}, 095023 (2012)
  [arXiv:1202.2014 [hep-ph]].
  %%CITATION = ARXIV:1202.2014;%%

%\cite{Englert:2013zpa}
\bibitem{Englert:2013zpa} 
  C.~Englert, E.~Re and M.~Spannowsky,
  %``Triplet Higgs boson collider phenomenology after the LHC,''
  Phys.\ Rev.\ D {\bf 87}, 095014 (2013)
  [arXiv:1302.6505 [hep-ph]].
  %%CITATION = ARXIV:1302.6505;%%
  
%\cite{Chiang:2014bia}
\bibitem{Chiang:2014bia} 
  C.~W.~Chiang, S.~Kanemura and K.~Yagyu,
  %``Novel Constraint on Parameter Space of the Georgi-Machacek Model by Current LHC Data,''
  Phys.\ Rev.\ D {\bf 90}, 115025 (2014)
  [arXiv:1407.5053 [hep-ph]].
  %%CITATION = ARXIV:1407.5053;%%
    
\bibitem{Khachatryan:2014jba} 
  V.~Khachatryan {\it et al.}  [CMS Collaboration],
  %``Precise determination of the mass of the Higgs boson and tests of compatibility of its couplings with the standard model predictions using proton collisions at 7 and 8 TeV,''
  arXiv:1412.8662 [hep-ex].
  %%CITATION = ARXIV:1412.8662;%%

\bibitem{ATLAS2015} 
  The ATLAS collaboration,
  %``Measurements of the Higgs boson production and decay rates and coupling strengths using pp collision data at √s = 7 and 8 TeV in the ATLAS experiment,''
  ATLAS-CONF-2015-007, ATLAS-COM-CONF-2015-011.
  %%CITATION = ATLAS-CONF-2015-007, ATLAS-COM-CONF-2015-011;%%

\bibitem{ATLAS2014} 
  The ATLAS collaboration,
  %``Evidence for electroweak production of $W^{\pm}W^{\pm}jj$ in $pp$ collisions at $\sqrt{s}=8$ TeV with the ATLAS detector,''
  ATLAS-CONF-2014-013, ATLAS-COM-CONF-2014-015.
  %%CITATION = ATLAS-CONF-2014-013, ATLAS-COM-CONF-2014-015;%%

\bibitem{Chiang:2015kka} 
  C.~W.~Chiang and K.~Tsumura,
  %``Properties and searches of the exotic neutral Higgs bosons in the Georgi-Machacek model,''
  arXiv:1501.04257 [hep-ph].
  %%CITATION = ARXIV:1501.04257;%%


  
\end{thebibliography}

\end{document}